\documentclass[11pt,authoryear]{article} 
\usepackage[paperheight=11.7in,paperwidth=8.3in]{geometry}

\usepackage[authoryear,round]{natbib}
\usepackage{latexsym,amssymb,amsmath}
\usepackage{longtable}
\usepackage{tikz,pgf}
\usepackage{subfig}
\usepackage{wrapfig}
\usepackage{rotating}

\newtheorem{theorem}{\bf Theorem}

\newtheorem{example}{\bf Example}

\newtheorem{definition}{Definition}

\newcommand{\bq}{\begin{eqnarray*}}
\newcommand{\eq}{\end{eqnarray*}}               
\newcommand{\bqn}{\begin{eqnarray}}
\newcommand{\eqn}{\end{eqnarray}}

\usepackage{setspace}
\usepackage{graphicx}
\usepackage{epstopdf}

\usepackage{url}

\begin{document}

\title{Topological Brain Network Distances}

\author{
  \begin{minipage}{\linewidth}
 \begin{center}
    Moo K. Chung$^{1}$
,Hyekyoung Lee$^{2}$, Andrey Gritsenko$^{1}$, \\ Alex DiChristofano$^{3}$,  Dustin Pluta$^{4}$, 
Hernando Ombao$^{4}$, Victor Solo$^{5}$\\
  \vspace{7pt}
      \textnormal {
$^{1}${University of Wisconsin, Madison, USA}\\
$^{2}${Seoul National University, Seoul, Korea}\\
$^{3}${University of Washington, St. Louis, USA}\\
$^{4}${King Abdullah University of Science and Technology, Saudi Arabia}\\
$^{5}${University of New South Wales, Sydney, Australia}\\
\vspace{7pt}
        {\tt mkchung@wisc.edu
        }\\
      }
      \end{center}
  \end{minipage}\\
}

\maketitle


\begin{abstract}
Existing brain network distances are often based on matrix norms. The element-wise differences in the existing matrix norms may fail to capture underlying topological differences. Further, matrix norms are sensitive to outliers. A major disadvantage to element-wise distance calculations is that it could be severely affected even by a small number of extreme edge weights. 
Thus it is necessary to develop network distances that recognize topology. In this paper, we provide a survey of bottleneck, Gromov-Hausdorff (GH) and Kolmogorov-Smirnov (KS) distances that are adapted for brain networks, and compare them against matrix-norm based network distances. Bottleneck and GH-distances are often used in persistent homology. However, they were rarely utilized to measure similarity between brain networks. 
KS-distance is recently introduced to measure the similarity between networks across different filtration values. The performance analysis was conducted using the random network simulations with the ground truths. Using a twin imaging study, which provides biological ground truth, we demonstrate that the KS distance has the ability to determine heritability.

\end{abstract}


\section{Introduction}

There are many similarity measures and distances between networks in literature \citep{banks.1994,chung.2013.MICCAI,lee.2012.TMI, chen.2016,petri.2014}. Many of these approaches simply ignore the topology of the networks and mainly use the sum of element-wise differences between either node or edge measurements. These network distances are sensitive to the topology of networks. 
One serious limitation of these approaches is that it could be insensitive to topological structures such as connected components, modules and holes in networks which are all essential to studying topological properties of networks.

In standard graph theoretic approaches,  the similarity and distance of networks are measured mainly by determining the difference in graph theory features such as assortativity, betweenness centrality, small-worldness and network homogeneity \citep{bullmore.2009,rubinov.2009.HBM,rubinov.2010.NI,uddin.2008}. Comparison of graph theory features appears to reveal changes of structural or functional connectivity associated with different clinical populations \citep{rubinov.2010.NI}. Since weighted brain networks are difficult to interpret and visualize, they are often turned into binary networks by thresholding edge weights \citep{he.2008,vanwijk.2010}. However, the thresholds for the edge weights are often chosen arbitrarily and produce results that could alter the network topology and thus make comparisons difficult. To obtain the proper optimal threshold where comparisons can be made, the multiple comparison correction over every possible edge has been proposed \citep{rubinov.2009.HBM,vanwijk.2010}. However, the resulting binary graph is extremely sensitive depending on the chosen $p$-value threshold value. Others tried to control the sparsity of edges in the network in obtaining the binary network \citep{achard.2007,bassett.2006,he.2008,vanwijk.2010,lee.2012.TMI}. However, one encounters the problem of thresholding sparse parameters. Thus existing methods for binarizing weighted networks cannot escape the inherent problem of arbitrary thresholding. 

There is currently no widely accepted criteria for thresholding networks. Instead of trying to find an optimal threshold that gives rise to a single network that may not be suitable for comparing clinical populations, cognitive conditions or different studies, {\em why not use each network produced from every threshold?} Motivated by this simple question, new multiscale hierarchical network modeling framework based on persistent homology has been proposed  \citep{chung.2013.MICCAI,lee.2011.MICCAI,lee.2011.ISBI,lee.2012.TMI}. Persistent homology, a branch of computational topology \citep{carlsson.2008,edelsbrunner.2008}, provides a more coherent mathematical framework to measuring network distance than the conventional method of simply taking difference between graph theoretic features or norm of the connectivity matrices. Instead of looking at networks at a fixed scale, as usually done in many standard brain network analysis, persistent homology observes the changes of topological features of the network over multiple resolutions and scales \citep{edelsbrunner.2008,horak.2009,zomorodian.2005}. In doing so, it reveals the most persistent topological features that are robust under noise perturbations. This robustness in performance under different scales is needed for most network distances that are parameter and scale dependent.  

In persistent homology based brain network analysis,  instead of analyzing networks at one fixed threshold that may not be optimal, we build the collection of nested networks over every possible threshold using the {\em graph filtration}, a persistent homological construct  \citep{lee.2011.MICCAI, lee.2012.TMI, chung.2013.MICCAI}. The graph filtration is a threshold-free framework for analyzing a family of graphs but requires hierarchically building specific nested subgraph structures. The graph filtration shares similarities to the existing multi-thresholding or multi-resolution network models  that use many different arbitrary thresholds or scales  \citep{achard.2006,he.2008,lee.2012.TMI,kim.2015,supekar.2008}. Such approaches are mainly used to visually display the dynamic pattern of how graph theoretic features change over different thresholds and the pattern of change is rarely quantified. Persistent homology can be used to quantify such dynamic pattern in a more coherent mathematical framework.

In persistent homology, there are various metrics that have been proposed to measure topological distances. Among them, {\em bottleneck distance} and {\em Gromov-Hausdorff (GH) distance} are possibly the two most popular distances that were originally used to measure distance between two metric spaces \citep{tuzhilin.2016}. They were later adapted to measure distances in persistent homology, dendrograms \citep{carlsson.2008,carlsson.2010,chazal.2009} and brain networks \citep{lee.2011.MICCAI,lee.2012.TMI}.  The probability distributions of bottleneck and GH-distances are unknown. Thus,  the statistical inference on them can only be done through resampling techniques such as 
permutations \citep{lee.2012.TMI,lee.2017.HBM}, which often cause serious computational bottlenecks for large-scale networks. To bypass the computational bottleneck associated with resampling large-scale networks, the {\em Kolmogorov-Smirnov (KS) distance} was introduced \citep{chung.2013.MICCAI,chung.2017.IPMI,lee.2017.HBM}. The advantage of using KS-distance is that its gives results that are easier to interpret than those obtained from less intuitive distances from persistent homology. Further, due to its simplicity in construction, it is possible to determine its probability distribution exactly without resampling \citep{chung.2017.IPMI}. 

The paper is organized into sections introducing matrix norm based distances and topological distances. 
Then each distance will be compared against each other in various random network simulation settings with the known ground truth. The method is applied to twin imaging data, where the biological ground truth is given by genetic difference observed in monozygtoic (MZ) and same-sex dizygotic (DZ) differences. 

\section{Matrix norm  and log-Euclidean distances}
\label{sec:network_construction}

Many distance or similarity measures are not metrics. However, having metric distances facilitate ease of interpretation of brain networks.  Further, existing network distance concepts are often borrowed from  the metric space theory. Let us start with formulating brain networks as metric spaces. 

Consider a weighted graph or network with the node set $V = \left\{ 1, \dots, p \right\}$ and the edge weights $w=(w_{ij})$, where $w_{ij}$ is the weight between nodes $i$ and $j$.
The edge weight is usually given by a similarity measure between the observed data on the nodes. Various similarity measures  have been proposed. The correlation or mutual information between measurements for the biological or metabolic network and the frequency of contact between actors for the social network have been used as edge weights \citep{li.2009,mcintosh.1994,newman.1999,song.2005,bassett.2006.adaptive,bien.2011}. We may assume that the edge weights satisfy the metric properties: nonnegativity, identity, symmetry and the triangle inequality such that
$$w_{i,j} \geq 0, \; w_{ii} =0,\; w_{ij} = w_{ji}, \;w_{ij} \leq w_{ik} + w_{kj}.$$ 
With theses conditions, $\mathcal{X}=(V, w)$ forms a metric space. Although the metric property is not necessary for building a network, it offers many nice mathematical properties and easier interpretation on network connectivity.

\begin{example}
\label{ex:corr}
Given measurement vector ${\bf x}_i  = (x_{1i}, \cdots, x_{ni})^{\top} \in \mathbb{R}^n$ on the node $i$. 
The weight $w=(w_{ij})$ between nodes is often given by some bivariate function $f$: $w_{ij} = f({\bf x}_i, {\bf x}_j)$. The Pearson correlation between ${\bf x}_i$ and ${\bf x}_j$, denoted as $\mbox{corr}({\bf x}_i, {\bf x}_j)$, is a bivariate function. If the weights $w=(w_{ij})$ are given by $w_{ij} = \sqrt{1-\mbox{corr}({\bf x}_i, {\bf x}_j)}$,  it can be shown that $\mathcal{X}=(V, w)$ forms a metric space.
\end{example}

Matrix norm of the difference between networks is often used as a measure of similarity between networks  \citep{banks.1994,zhu.2014}. Given two networks $\mathcal{X}^1=(V, w^1)$ and $\mathcal{X}^2=(V, w^2)$, the $L_l$-norm of network difference is given by
$$D_l (\mathcal{X}^1,\mathcal{X}^2) = \parallel w^1 - w^2 \parallel_{l} =
            \Big(  \sum_{i,j}  \big| w^1_{ij} - w^2_{ij} \big|^{l}  \Big)^{1/l}. 
\label{eq:norm}$$
Note $L_l$ is the element-wise Euclidean distance in $l$-dimension. When $l=\infty,$ $L_{\infty}$-distance is written as 
$$
 D_{\infty} (\mathcal{X}^1,\mathcal{X}^2) = \parallel w^1 - w^2 \parallel_{\infty} =
            \max_{\forall i,j}  \big| w^1_{ij} - w^2_{ij} \big|. 
\label{eq:inf_norm}
$$
The element-wise differences  may not capture additional higher order similarity. For example, it may be insensitive to relationships between pairs of columns or rows \citep{zhu.2014}. Also, one major disadvantage of $L_1$- and $L_2$-distances is that they are not robust against outliers. Few outlying extreme edge weights may severely affect the distance. Further, these distances ignore the underlying topological structures. This paper highlights the need to develop distances that respect the topological structure of networks.

\section{Correlation metric space}
We now show how to construct a class of metric spaces using correlations. Consider an $n \times 1$ measurement vector ${\bf x}_j = (x_{1j},\cdots,$ $x_{nj})^\top$ on node $j$ that is standardized to have mean 0 and norm 1 such that
\bqn  \parallel {\bf x}_j \parallel^2 = {\bf x}_{j}' {\bf x}_{j} = \sum_{i=1}^n x_{ij}^2 = 1, \quad \sum_{i=1}^n x_{ij} = 0. \eqn
Then ${\bf x}_i^{\top} {\bf x}_j$ is the Pearson correlation between ${\bf x}_i$ and ${\bf x}_j$. Note that correlations are invariant under scale and translations. Naturally, we are interested in using correlations or their simple functions as edge wights such as  
$$\rho_{ij} = {\bf x}_i^\top{\bf x}_j \quad \mbox{ or  } \quad \rho_{ij} = 1- {\bf x}_i^\top{\bf x}_j.$$ 
However, not every functions of correlations are metric. 

\begin{example} 
$\rho_{ij} = 1- {\bf x}_i^\top{\bf x}_j$ is {\em not} metric. 
Consider the following 3-node counter example: 
\bqn {\bf x}_i &=& (0, \frac{1}{\sqrt{2}},  -\frac{1}{\sqrt{2}})^\top,\\
{\bf x}_j &=& (\frac{1}{\sqrt{2}},  0, -\frac{1}{\sqrt{2}})^\top,\\  
{\bf x}_k &=& (\frac{1}{\sqrt{6}}, \frac{1}{\sqrt{6}}, -\frac{2}{\sqrt{6}})^\top.
\eqn
Then we have $\rho_{ij} > \rho_{ik} + \rho_{jk}$. 
\end{example}

Then an interesting question is to identify minimum conditions that make a function of correlations a metric.

\begin{theorem}For centered and scaled data ${\bf x}_1, \cdots, {\bf x}_p$, let $$\rho_{ij} = cos^{-1}({\bf x}_i^\top{\bf x}_j).$$ Then $\rho_{ij}$ is metric. 
\label{theorem:metric}
\end{theorem}

{\em Proof.}
On unit sphere $S^{n-1}$, the correlation between ${\bf x}_i$ and ${\bf x}_j$ is given by the cosine angle $\theta_{ij}$ between the two vectors, i.e., $${\bf x}_i^\top {\bf x}_j = \cos \theta_{ij}.$$
The geodesic distance $\rho$ between nodes ${\bf x}_i$ and ${\bf x}_j$ on the unit sphere is given by angle $\theta_{ij}$:
$$\rho( {\bf x}_i, {\bf x}_j) = \cos^{-1} ({\bf x}_i^\top {\bf x}_j).$$ 
For nodes ${\bf x}_i, {\bf x}_j \in S^{n-1}$, there are two possible angles $\theta_{ij}$ and $2\pi - \theta_{ij}$ depending on if we measure the angles along the shortest arc or longest arc. We take the convention of using the smallest angle in defining $\theta_{ij}$. 
With this convention, 
$$0 \leq \rho( {\bf x}_i, {\bf x}_j)  \leq \pi.$$
Given three nodes ${\bf x}_i, {\bf x}_j$ and ${\bf x}_k$, which forms a spherical triangle, 
we then have spherical triangle inequality
\bq \rho( {\bf x}_i, {\bf x}_j) \leq \rho( {\bf x}_i, {\bf x}_k) + \rho( {\bf x}_k, {\bf x}_j). \label{eq:STI} \eq
The proof to (\ref{eq:STI}) is given in \citet{reid.2005}. Thus we proved $\rho$ is a metric. $\square$

\begin{theorem} For any metric $\rho_{ij}$, $f(\rho_{ij})$ is also a metric if $f(0)=0$ and $f(x)$ is increasing and concave for $x >0$
\end{theorem}
The proof is given in \citet{van.2012}. Such function $f$ is called the {\em metric preserving function}. Any power $[cos^{-1}({\bf x}_i^\top{\bf x}_j)]^{1/m}$ for $m \geq 1$ is metric. When $m=1$, we have the simplest possible metric 
$\rho({\bf x}_i, {\bf x}_j) = \cos^{-1}({\bf x}_i^\top{\bf x}_j)$,
which obtains minimum $0$ when ${\bf x}_i^\top{\bf x}_j = 1$ and maximum $\pi$ when ${\bf x}_i^\top{\bf x}_j = -1$. From \label{ex:corr}, since $w_{ij} = \sqrt{ 1- corr({\bf x}_i, {\bf x}_j)}$ is metric, any power $[ 1- corr({\bf x}_i, {\bf x}_j) ]^{1/2m}$ for $m \geq 1$ is also metric.

\section{Graph filtration}

All topological network distances that will be introduced in later sections are based on filtrations on graphs by thresholding edge weights. 

\begin{definition} Given weighted network $\mathcal{X}=(V, w)$ with edge weight $w = (w_{ij})$, the binary network $\mathcal{X}_{\epsilon} =(V, w_{\epsilon})$ is a graph consisting of the node set $V$ and the binary edge weights 
$w_{\epsilon}$ given by 
\bqn w_{\epsilon} = (w_{ij,{\epsilon}}) =   \begin{cases}
1 &\; \mbox{  if } w_{ij} > \epsilon;\\
0 & \; \mbox{ otherwise}.
\end{cases}
\label{eq:case}
\eqn
\end{definition}
Any edge weight less than or equal to $\lambda$ is made into zero while edge weight larger than $\lambda$ is made into one. Note \citet{lee.2012.TMI} defines the binary graphs by thresholding above, which is consistent with the definition of the Rips filtration. However, in brain imaging, higher value  of $w_{ij}$ indicates stronger connectivity. Thus, we are thresholding below and leave out stronger connections \citep{chung.2013.MICCAI}.

Note $w_{\epsilon}$ is the adjacency matrix of $\mathcal{X}_{\epsilon}$, which is a simplicial complex consisting of $0$-simplices (nodes) and $1$-simplices (edges)  \citep{ghrist.2008}. 
In the metric space $\mathcal{X}=(V, w)$, the Rips complex $\mathcal{R}_{\epsilon}(X)$ is a simplicial complex whose $(p-1)$-simplices correspond to unordered $p$-tuples of points that satisfy $w_{ij} \leq \epsilon$ in a pairwise fashion \citep{ghrist.2008}. While the binary network $\mathcal{X}_{\epsilon}$ has at most 1-simplices, the Rips complex can have at most $(p-1)$-simplices. Thus, the compliment of the binary graph, which thresholds above, satisfies $\mathcal{X}_{\epsilon}^c \subset \mathcal{R}_{\epsilon}(\mathcal{X})$. The Rips complex has the property that
$$\mathcal{R}_{\epsilon_0}(\mathcal{X}) \subset \mathcal{R}_{\epsilon_1}(\mathcal{X}) \subset \mathcal{R}_{\epsilon_2}(\mathcal{X}) \subset \cdots $$
for $0=\epsilon_{0} < \epsilon_{1} < \epsilon_{2} < \cdots.$
When $\epsilon=0$, the Rips complex is simply the node set $V$. By increasing the filtration value $\epsilon$, we are connecting more nodes so the size of the edge set increases.
Such the nested sequence of the Rips complexes is called a Rips filtration, the main object of interest in the persistent homology \citep{edelsbrunner.2008}.

Since a binary network is a special case of the Rips complex, we have
$$\mathcal{X}_{\epsilon_0}^c  \subset \mathcal{X}_{\epsilon_1}^c  \subset \mathcal{X}_{\epsilon_2}^c \subset \cdots.$$
Equivalently, we also have
$$\mathcal{X}_{\epsilon_0}  \supset \mathcal{X}_{\epsilon_1}  \supset \mathcal{X}_{\epsilon_2} \supset \cdots.$$
The sequence of such nested multiscale graphs  is defined as the {\em graph filtration} \citep{lee.2011.MICCAI,lee.2012.TMI}. 
Note that $\mathcal{X}_0$ is the complete weighted graph while $\mathcal{X}_{\infty}$ is the node set $V$. By increasing the threshold value, we are thresholding at higher connectivity so more edges are removed. Given a weighted graph, there are infinitely many different filtrations. For different $\epsilon_j$ and $\epsilon_{j+1}$, we can have identical binary graph, i.e., $\mathcal{X}_{\epsilon_j}  = \mathcal{X}_{\epsilon_{j+1}}$. So a question naturally arises if there is a unique filtration that can be used in practice.  Let the {\em level of a filtration} be the number of nested unique sublevel sets in the given filtration. 

\begin{theorem} 
\label{theorem:maximal}
For graph $X=(V, w)$ with $q$ unique positive edge weights, the maximum level of a filtration  is $q+1$. Further, the  filtration with $q+1$ filtration level is unique.
\end{theorem}
{\em Proof.} For a graph with $p$ nodes, the maximum number of edges is $(p^2-p)/2$, which is obtained in a complete graph. If we order the edge weights in the increasing order, we have the sorted edge weights:
$$0 = w_{(0)} <  \min_{j,k} w_{jk} = w_{(1)} < w_{(2)} < \cdots < w_{(q)} = \max_{j,k} w_{jk},$$
where $q \leq (p^2-p)/2$.  The subscript $_{( \;)}$ denotes the order statistic. 
For all $\lambda < w_{(1)}$, $\mathcal{X}_{\lambda} = \mathcal{X}_{0}$ is the complete graph of $V$. For all $w_{(r)}  \leq \lambda < w_{(r+1)} \; (r =1, \cdots, q-1)$, $\mathcal{X}_{\lambda} = \mathcal{X}_{w_{(r)}}$. For all $ w_{(q)} \leq \lambda$, $\mathcal{X}_{\lambda}= \mathcal{X}_{\rho_{(q)}} =V$, the vertex set. Hence, the filtration given by
\bq  \mathcal{X}_{0}  \supset  \mathcal{X}_{w_{(1)}}  \supset  \mathcal{X}_{w_{(2)}}  \supset \cdots  \supset  \mathcal{X}_{w_{(q)}}\label{eq:maximal}\eq
is {\em maximal} in a sense that we cannot have any additional level of filtration. $\square$

Through the paper, the {\em maximal graph filtration} (\ref{eq:maximal}) will be used. The condition of having unique edge weights in Theorem \ref{theorem:maximal} is not restrictive in practice. Assuming edge weights to follow some continuous distribution, the probability of any two edges being equal is zero. Among many possible filtrations, we will use the maximal filtration (\ref{eq:maximal}) in the study since it is uniquely given. The finiteness and uniqueness of the filtration levels over finite graphs are intuitively clear by themselves and are implicitly assumed in software packages such as javaPlex \citep{adams.2014}. However, we still needed a rigorous statement to specify the type of filtration we are using.

On the maximal graph filtration, $\beta_0$ and $\beta_1$ numbers have very stable monotonic increase and decrease.  

\begin{theorem} In a graph, Betti numbers $\beta_0$ and $\beta_1$ are monotone over filtration on edge weights.
\end{theorem}
{\em Proof.} 
Under graph filtration in (\ref{eq:maximal}), the edges are deleted one at a time. Since an edge has only two end points, the deletion of the edge disconnect the graph into at most two. Thus, the number of connected components ($\beta_0$) always increases and the increase is at most by one. The Euler characteristic $\chi$ of the graph is given by \citep{adler.2010}
$$\chi = \beta_0 - \beta_1  = p - q,$$
where $p$ and $q$ are the number of nodes and edges respectively. Thus,
$$\beta_1 = \beta_0 - p + q.$$
Note $p$ is fixed over the filtration but $r$ is decreasing by one while $\beta_0$ increases at most by one. Hence, $\beta_1$ always decreases and the decrease is at most by one. $\square$

Once we compute $\beta_0$ numbers, $\beta_1$ number is simply given by $\beta_0 - p + q$.

\section{Bottleneck distance} 
This is perhaps the most often used distance in persistent homology but it is rarely used for brain networks. In persistent homology, the topology of underlying data can be represented by the birth and death of holes. In a graph, the 0- and 1-dimensional holes are a connected component and a cycle, respectively \citep{carlsson.2008.JCV}. During a filtration, holes in a homology group appear and disappear. The Betti number at a particular threshold is then the number of holes at that threshold. If a  hole appears at the threshold $\xi$ and disappears at $\tau,$ it can be encoded into a point, $(\xi,\tau)  \; (0 \le \xi \le \tau < \infty)$ in $\mathbb{R}^2$. If $m$ number of holes appear during the filtration of a network $\mathcal{X}=(V,w)$, the homology group can be represented by a point set 
$$\mathcal{P} (\mathcal{X}) = \left\{ (\xi_{1},\tau_{1}), \dots, (\xi_{m},\tau_{m}) \right\}.$$ 
This scatter plot is called the persistence diagram (PD) \citep{cohensteiner.2007}. 

\begin{figure}[th!]
\centering
\includegraphics[width=1\linewidth]{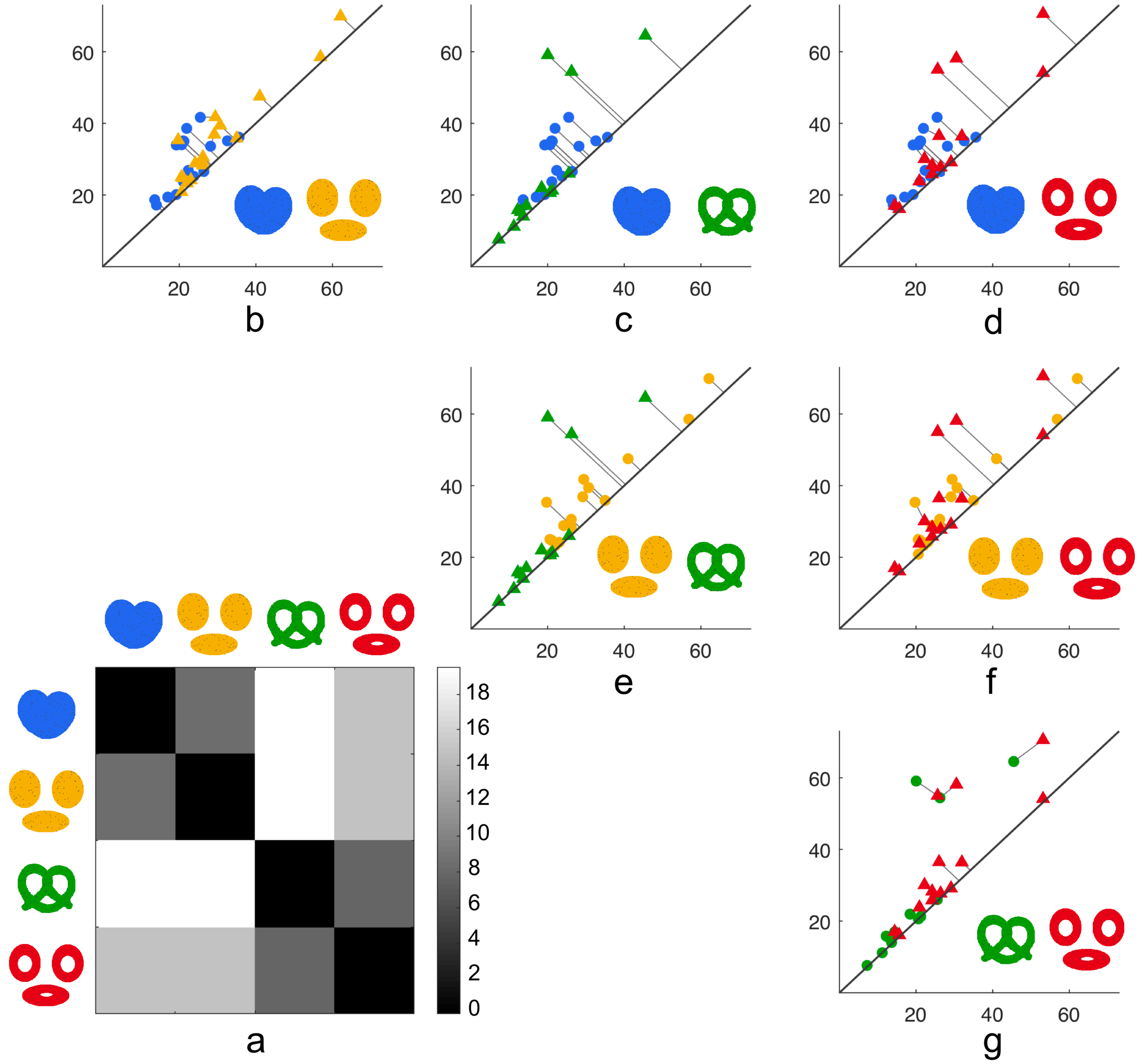}
\caption{Toy example of bottleneck distance on four different types of scatter points: $\mathcal{X}^1$ (blue), $\mathcal{X}^2$ (yellow), $\mathcal{X}^3$ (green), $\mathcal{X}^4$ (red). (a) Bottleneck distance displayed as  a connectivity matrix (b-g) Persistence diagrams of holes of the pair of topological spaces. Blue, yellow, green, and red points correspond to the holes of blue, yellow, green, and red topological spaces, respectively. The projection line to $y=x$ indicates the argumented points.}
\label{fig:example_bottleneck}
\end{figure}

Given two networks $\mathcal{X}^1=(V^1, w^1)$ with $m$ holes and $\mathcal{X}^2=(V^2, w^2)$ with $n$ holes,  we construct the corresponding graph filtrations. Subsequently, PDs $$\mathcal{P} (\mathcal{X}^1) = \left\{ (\xi_{1}^{1},\tau_{1}^{1}), \cdots, (\xi_{m}^{1},\tau_{m}^{1}) \right\}$$ and $$\mathcal{P} (\mathcal{X}^2) = \left\{ (\xi_{1}^{2},\tau_{1}^{2}), \cdots, (\xi_{n}^{2},\tau_{n}^{2}) \right\}$$ are obtained through the filtration \citep{lee.2012.TMI}. The bottleneck distance between the networks is defined as the bottleneck distance of the corresponding PDs \citep{cohensteiner.2007}:
\bq
D_{B} \big(\mathcal{P} (\mathcal{X}^1),\mathcal{P} (\mathcal{X}^2) \big) = 
\inf_{\gamma} \sup_{1 \leq i \leq m}  \parallel t_i^1 - \gamma(t_i^1) \parallel_{\infty},
\label{eq:D_B}
\eq
where $t_i^1 =(\xi_{i}^{1},\tau_{i}^{1})  \in \mathcal{P} (\mathcal{X}^1)$ and $\gamma$ is a bijection from $\mathcal{P} (\mathcal{X}^1)$ to $\mathcal{P} (\mathcal{X}^2)$. The infimum is taken over all possible bijections. If  $t_{j}^{2} = (\xi_{j}^{2},\tau_{j}^{2}) = \gamma(t_{i}^{1})$ for some $i$ and $j$, $L_{\infty}$-norm is given by
$$\parallel t_{i}^{1} - \gamma(t_{i}^{1}) \parallel_{\infty} = \max \big( | \xi_{i}^{1}-\xi_{j}^{2}|,| \tau_{i}^{1}-\tau_{j}^{2}| \big).$$ 
Note (\ref{eq:D_B}) assumes $m=n$ such that the bijection $\gamma$ exists. Suppose two networks share the same node set, i.e., $V^1 = V^2$, with $p$ nodes and  the same number of $q$ unique edge weights. If  the maximal
filtration is performed on two networks, the number of their 0- and 1-dimensional holes that appear and disappear during the filtration is $p$ and $1-p+q$, respectively. Thus, their persistence diagrams of 0- or 1-dimensional holes always have the same number of points. The bijection  $\gamma$ is determined by the bipartite graph matching algorithm \citep{cohensteiner.2007,edelsbrunner.2008}.

If $m \neq n$, there is no one-to-one correspondence between two PDs. Then, auxiliary points  $$(\frac{\xi_{1}^{1}+\tau_{1}^{1}}{2},\frac{\xi_{1}^{1}+\tau_{1}^{1}}{2}), \cdots, (\frac{\xi_{m}^{1}+\tau_{m}^{1}}{2},\frac{\xi_{m}^{1}+\tau_{m}^{1}}{2})$$ and $$(\frac{\xi_{1}^{2}+\tau_{1}^{2}}{2},\frac{\xi_{1}^{2}+\tau_{1}^{2}}{2}), \cdots, (\frac{\xi_{n}^{2}+\tau_{n}^{2}}{2},\frac{\xi_{n}^{2}+\tau_{n}^{2}}{2})$$ that are orthogonal projections to  the diagonal line $\xi = \tau$ in $\mathcal{P} (\mathcal{X}^1)$ and $\mathcal{P} (\mathcal{X}^2)$ are added to  $\mathcal{P} (\mathcal{X}^2)$ and $\mathcal{P} (\mathcal{X}^1)$ respectively to make the identical number of points in PDs (Figure \ref{fig:example_bottleneck}).


\begin{example}
\label{ex:bottleneck}
The bottleneck distances between four toy examples are displayed in Fig. \ref{fig:example_bottleneck}.  
There are four topologically different scatter point data $\mathcal{X}^1, \mathcal{X}^2, \mathcal{X}^3, $ and $\mathcal{X}^4$ that are represented by blue, yellow, green, and red in (a).
Their pairwise PDs of $\beta_1$ are shown in (b-g). 
In terms of $\beta_1$, it can easily distinguish $\mathcal{X}^1 in blue, \mathcal{X}^2$ in yellow (no hole) from $\mathcal{X}^3 in green, \mathcal{X}^4$ in red (three holes).
\end{example}

\section{Gromov-Hausdorff distance} 
\label{sec:gh_distance}

\begin{figure}[t]
\centering
\includegraphics[width=1\linewidth]{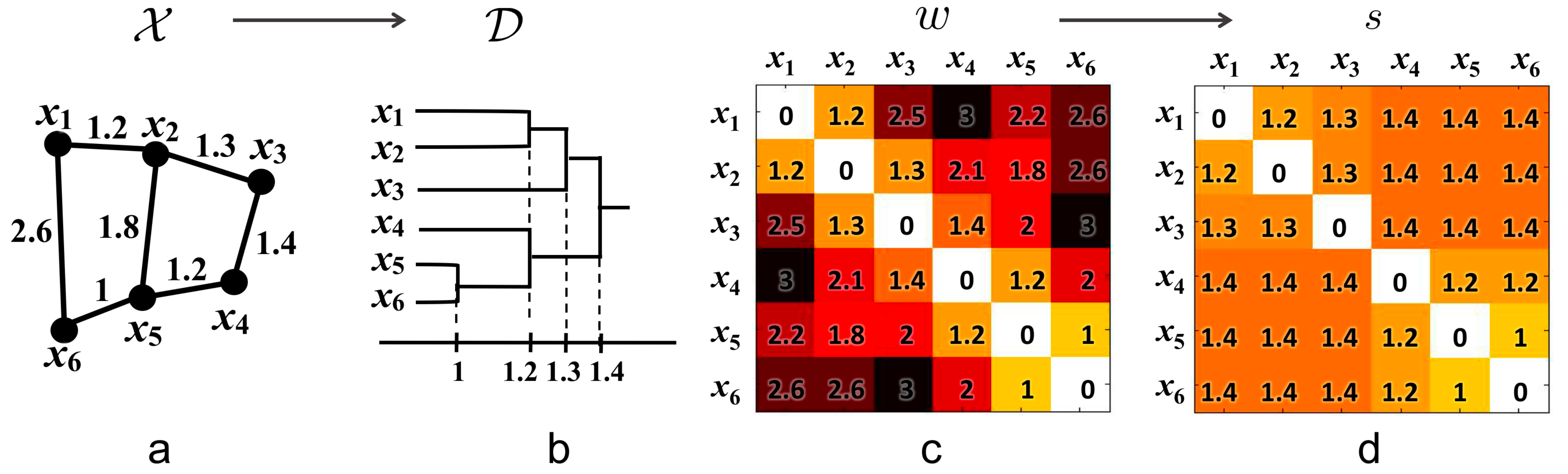}
\caption{(a) Toy network, (b) its dendrogram, (c) the distance matrix $w$ based on Euclidean distance, (d) the single linkage matrix $S$.}
\label{fig:example_GHdistance}
\end{figure}

Gromov-Hausdorff (GH) distance has been used to characterize network distance in brain images \citep{lee.2011.MICCAI,lee.2012.TMI,chung.2017.CNI}. GH-distance measures the difference between networks by embedding the network into the ultrametric space that represents hierarchical clustering structure of network \citep{carlsson.2010}. The distance $s_{ij}$ between the closest nodes in the two disjoint connected components ${\bf R}_1$ and ${\bf R}_2$ is called the single linkage distance (SLD), which is defined as 
$$ s_{ij} = \min_{l \in {\bf R}_{1}, k \in {\bf R}_{2}} w_{lk}.$$ 
Every edge connecting a node in ${\bf R}_1$ to a node in ${\bf R}_2$ has the same SLD.
SLD is then used to construct the single linkage matrix (SLM) $S = (s_{ij})$ (Figure \ref{fig:example_GHdistance}).  
SLM shows how connected components are merged locally and can be used in constructing a dendrogram. SLM is a {\em ultrametric}
 which is a metric space satisfying the stronger triangle inequality $s_{ij} \le \max (s_{ik},s_{kj})$ \citep{carlsson.2010}.
Thus the dendrogram can be represented as a ultrametric space $\mathcal{D} = (V,S),$ which is again a metric space. 
GH-distance between networks is then defined through GH-distance between corresponding dendrograms. Given two dendrograms $\mathcal{D}^{1}=(V,S^{1})$ and $\mathcal{D}^{2}=(V,S^{2})$ with SLM $S^1 = (s^1_{ij})$ and $S^2 = (s^2_{ij})$,
\bqn D_{GH} (\mathcal{D}^1,\mathcal{D}^2) = \frac{1}{2} \max_{\forall i,j}  | s^1_{ij} - s^2_{ij} |. \label{eq:D_GH} \eqn
For the statistical inference on GH-distance, resampling techniques such as jackknife or permutation tests are often used \citep{
lee.2012.TMI,lee.2017.HBM}. In this study, we will use the permutation test.

\section{Permutation test on network distances}

Statistical inference on network distances can be done using the permutation test or bootstrap \citep{chung.2013.MICCAI,efron.1982,lee.2012.TMI}. Here we explain the permutation test procedure that was used for network distances. The usual setting in brain imaging applications is a two-sample comparison. 
Suppose there are $m$ measurement in Group 1 on node set $V$ of size $p$. Denote the data matrix as ${\bf X}^1_{m \times p}$. The edge weights of Group 1 is given by  $f({\bf X}^1)$  for some function $f$ and the metric space is given by $\mathcal{X}^1 = (V, f({\bf X}^1))$. Suppose there are $n$ measurement in Group 2 on the identical node set $V$. 
Denote data matrix as ${\bf X}^2_{n \times p}$ and the corresponding metric space as $\mathcal{X}^1 = (V, f({\bf X}^1))$.  We test the statistical significance of  network distance $D(\mathcal{X}^1, \mathcal{X}^2)$ under the null hypothesis $H_0$:
$$H_0: D(\mathcal{X}^1, \mathcal{X}^2)=0  \mbox{ vs. } H_1: D(\mathcal{X}^1, \mathcal{X}^2) > 0.$$

The permutation test is done as follows. Under $H_0$, one can concatenate the data matrices  
$${\bf X} = (x_{ij}) =
\left(
\begin{array}{c}
  {\bf X}^1     \\
  {\bf X}^2   \\
\end{array}
\right)_{(m+n) \times p}$$
and then permute the indices of the row vectors of ${\bf X}$ in the symmetric group of degree $m+n$, i.e., $S_{m+n}$ \citep{kondor.2007}. Denote the $i$-th permuted data matrix as ${\bf X}_{\sigma(i)} = (x_{\sigma(i), j})$, where $\sigma \in S_{m+n}$. Then we split ${\bf X}_{\sigma(i)}$ into submatrices such that
$${\bf X}_{\sigma(i)} =
\left(
\begin{array}{c}
  {\bf X}_{\sigma(i)} ^1     \\
  {\bf X}_{\sigma(i)} ^2   \\
\end{array}
\right),$$
where ${\bf X}_{\sigma(i)} ^1$ and  ${\bf X}_{\sigma(i)} ^2$ are of sizes $m \times p$ and $n \times p$ respectively.  Let $\mathcal{X}^1_{{\sigma}(i)} = (V, f({\bf X}^1_{{\sigma}(i)} ))$ and $\mathcal{X}^2_{{\sigma}(i)} = (V, f({\bf X}^2_{{\sigma}(i)} ))$ be weighted networks where the rows of the data matrices are permuted across the groups. Then we have distance $D(\mathcal{X}^1_{{\sigma}(i)} , \mathcal{X}^2_{{\sigma}(i)} )$ for each permutation. The fraction of permutations  $D(\mathcal{X}^1_{{\sigma}(i)} , \mathcal{X}^2_{{\sigma}(i)} )$ that is larger than $D(\mathcal{X}^1, \mathcal{X}^2)$ gives the estimate for $p$-value. The number of permutations exponentially increases and it is impractical to generate every possible permutation. So up to tens of thousands permutations are generated to guarantee convergence in practice. This is an approximate method and a care should be taken to guarantee the convergence but in most studies about 1\% of total permutations are used \citep{thompson.2001,zalesky.2010}.

\section{Kolmogorov-Smirnov distance} 

Recently Kolmogorov-Smirnov (KS) distance based on  graph filtration has been proposed and successfully applied to brain networks as a way to quantify networks without thresholding  \citep{chung.2017.IPMI}. The main advantage of the method is that it avoids using the computationally costly and time consuming permutation test for large-scale networks. 

The graph filtration can be quantified using monotonic function $f$ satisfying
\bqn f ( \mathcal{X}_{\epsilon_0} ) \geq f ( \mathcal{X}_{\epsilon_1} )  \geq f ( \mathcal{X}_{\epsilon_2} )  \geq \cdots   \label{eq:B} \eqn
or
\bqn f ( \mathcal{X}_{\epsilon_0} ) \leq f ( \mathcal{X}_{\epsilon_1} )  \leq f ( \mathcal{X}_{\epsilon_2} )  \leq \cdots   \label{eq:B} \eqn

The number of connected components (zero-th Betti number $\beta_0$) and the number of cycles (first Betti number $\beta_1$) satisfy the monotonicity. The size of the largest cluster (denoted as $\gamma$) satisfies a similar but opposite relation of monotonic increase. There are numerous  other monotone graph theory features that can be used \citep{chung.2013.MICCAI,chung.2017.IPMI}.

Given two networks $\mathcal{X}^1=(V, w^1)$ and $\mathcal{X}^2=(V, w^2)$, 
Kolmogorov-Smirnov (KS) distance between $\mathcal{X}^1$ and $\mathcal{X}^2$ is defined as \citep{chung.2013.MICCAI,lee.2017.HBM}
$$D_{KS}(\mathcal{X}^1, \mathcal{X}^2) = \sup_{\epsilon \geq 0} \big| f (\mathcal{X}^1_{\epsilon})  - f ( \mathcal{X}^2_{\epsilon}) \big|$$
using monotone function $f$. The distance $D_{KS}$ is motivated by Kolmogorov-Smirnov (KS) test for determining the equivalence of two cumulative distribution functions   \citep{bohm.2010, chung.2017.IPMI,gibbons.1992}. The distance $D_{KS}$ can be discretely approximated using the finite number of filtrations:
$$D_q = \sup_{1 \leq j \leq q} \big| f (\mathcal{X}^1_{\epsilon_j})  - f ( \mathcal{X}^2_{\epsilon_j}) \big|.$$
If we choose enough number of $q$ such that $\epsilon_j$ are all the sorted edge weights, then $$D_{KS}(\mathcal{X}^1,\mathcal{X}^2) = D_q$$ \citep{chung.2017.IPMI}. This is possible since there are only up to $p(p-1)/2$ number of unique edges in a graph with $p$ nodes and the monotone function increases discretely but {\em not continuously}. In practice,  $\epsilon_j$ may be chosen uniformly or a divide-and-conquer strategy can be used to do adaptively grid the filtration values.

\begin{theorem}
\label{thm:lim1}
$$P (D_q \geq d)   = 1 - \frac{A_{q,q}}{{2q \choose q}},$$
where $A_{u,v}$ satisfies $A_{u,v} = A_{u-1,v} + A_{u, v-1}$
with the boundary condition $A_{0,v}=A_{u,0}=1$ within band $|u - v| < d$ and initial condition $A_{0,0} =0$ for  $u,v \geq 1$.
\end{theorem} 
The proof is given in \citet{chung.2017.IPMI}. 
KS-distance method does not assume any statistical distribution on graph features other than that they has to be monotonic. The technique is very general and applicable to other monotonic graph features such as node degrees.

\begin{wrapfigure}{r}{0.4\textwidth}
\vspace{-1cm}
 \centering
\includegraphics[width=1\linewidth]{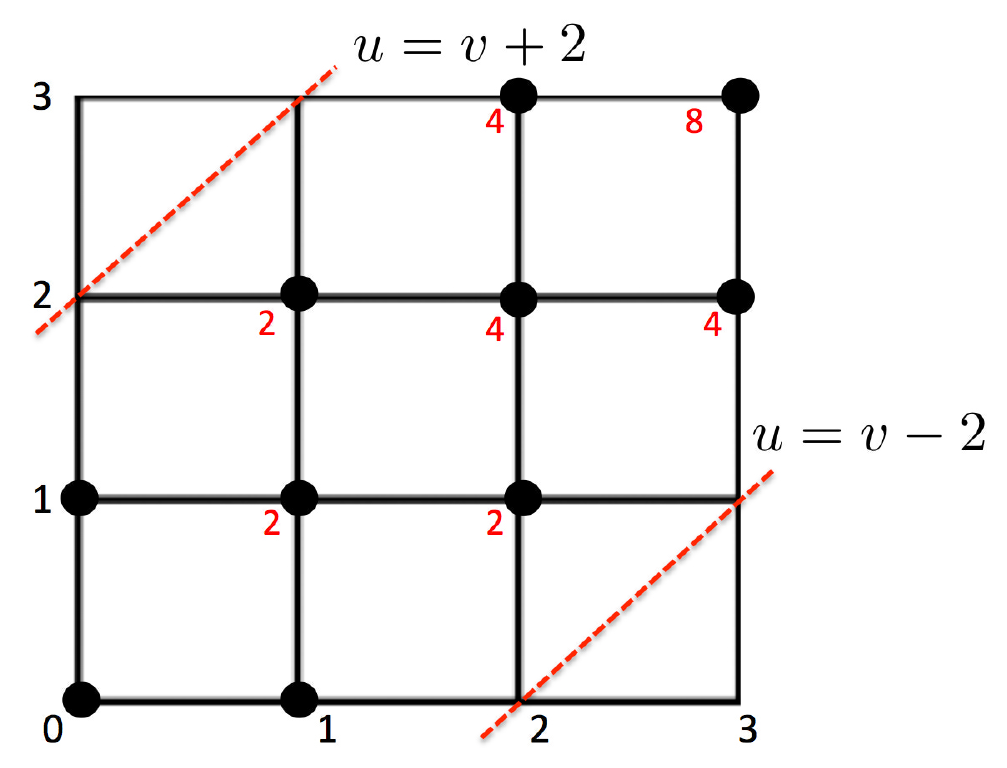} 
\caption{In this example, $A_{u,v}$ is computed within the boundary (dotted red line) from (0,0) to (3,3).}
\label{fig:App}
\vspace{-0.5cm}
\end{wrapfigure}

\begin{example}
$P(D_3 \geq 2)$ is computed sequentially as follows (Figure \ref{fig:App}).  We start with the bottom left corner $A_{0,0} = 0$ and move right or up toward the upper corner 
\bqn A_{1,0} &=& 1, A_{0,1} =1\\
 \to A_{1,1} &=& A_{1,0} + A_{0,1}\\ 
\to \cdots &=& \cdots\\
\to A_{3,3} &=& A_{3,2} + A_{2,3} =8.
 \eqn
 The probability is then $P(D_3 \geq 2) = 1- 8/ {6 \choose 3}=0.6$. The computational complexity of the combinatorial inference is $\mathcal{O}(q \log q)$ for sorting and $\mathcal{O}(q^2)$ for computing $A_{q,q}$ in the grid while the permutation test requires exponential run time.
\end{example}

When $q$ is too large, it may not be possible to represent and compute ${2q \choose q}$ in all the digits. For large $q$, use use the asymptotic probability distribution $D_q$ given by \citep{chung.2017.IPMI}
\bq \lim_{q \to \infty}  \Big( D_q /\sqrt{2q} \geq  d  \Big)  = 2 \sum_{i=1}^{\infty} (-1)^{i-1}e^{-2i^2d^2}. \label{eq:asympotic}\eq
$p$-value of the test statistic under the null is then computed as
\bqn \mbox{$p$-value} = 2 e^{-d_{o}^2} - 2e^{-8d_{o}^2} + 2 e^{-18d_{o}^2} \cdots, \label{eq:pvalue}\eqn
where the observed value $d_{o}$ is the  least integer greater than $D_{q}/\sqrt{2q}$ in the data.  For any large value $d_0 \geq 2$, the second term is in the order of $10^{-14}$ and insignificant. Even for small observed $d_0$, the expansion converges quickly and 5 terms are sufficient.

\section{Comparisons}

\begin{figure}[t]
\centering
\includegraphics[width=1\linewidth]{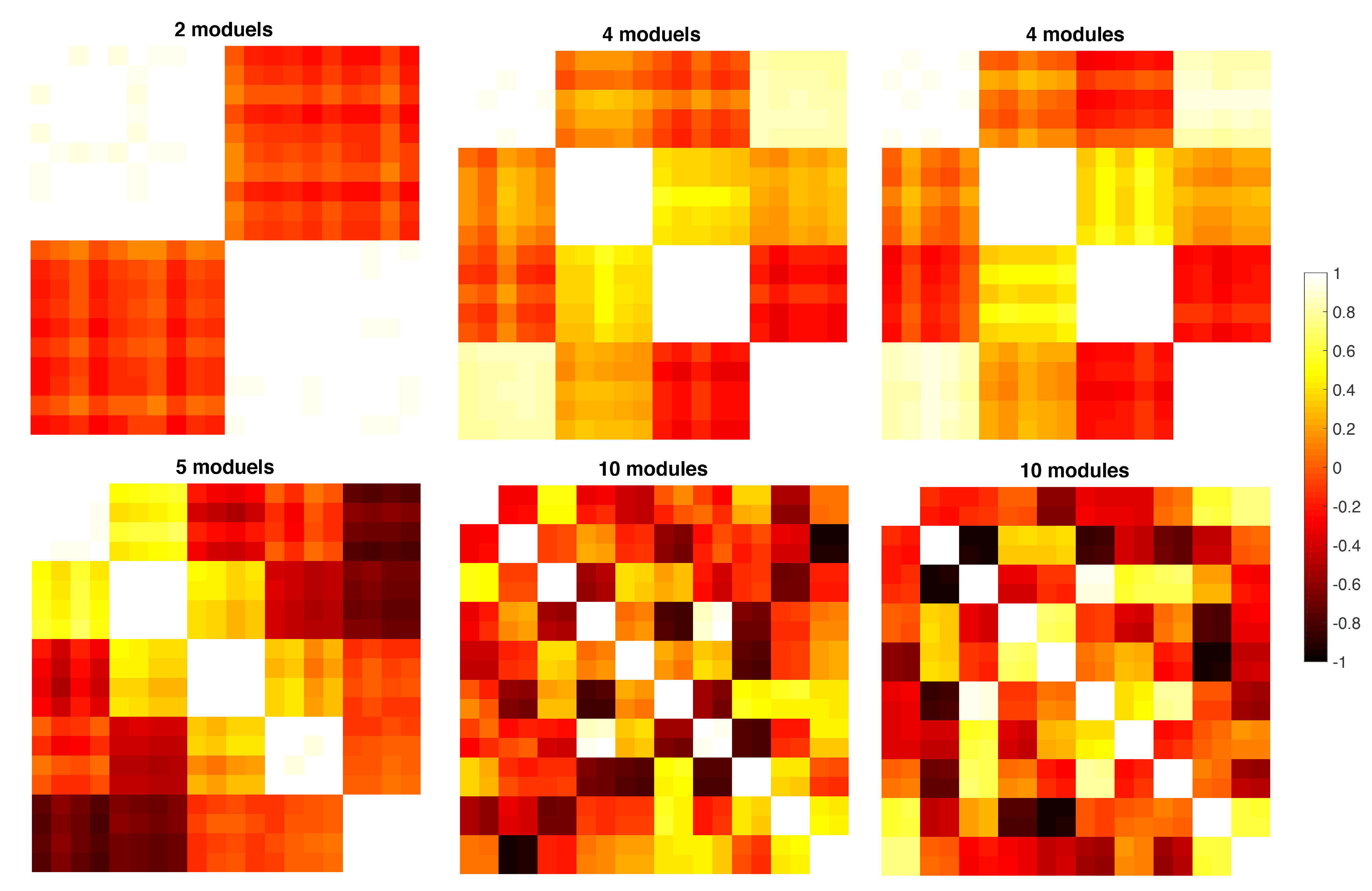}
\caption{Randomly simulated correlation matrices with modular structures used in the comparisons of network distances.}
\label{fig:simulation}
\end{figure}

Six different network distances ($L_1$, $ L_2$, $L_{\infty}$, Bottleneck, GH and KS) were compared in simulation studies. The simulations below were independently performed 100 times. We used $p=20$ nodes and  $n=5$ images in each group, which makes it possible for permutations to be exactly ${5 + 5 \choose 5} = 252$ (Figure \ref{fig:simulation}). 
The small number of permutation enables us to compare the performance of distance exactly. Through the simulations, $\sigma=0.1$ was universally used as network variability. 

The data vector ${\bf x}_i$ at node $i$ was simulated as identical and independently distributed multivariate normal across $i$, i.e., ${\bf x}_i \sim N(0,I_n)$ with $n$ by $n$ identity matrix $I_n$ as the covariance matrix. This gives the correlation matrix $C^1=(c_{ij}^1) = (corr({\bf x}_i, {\bf x}_j))$. The edge weights was given by $\sqrt{1- c^1_{ij}}$. The data vector ${\bf y}_i$ at node $i$ that produces modular structure was generated by adding additional dependency to ${\bf x}_i$ through a hierarchical  linear model \citep{snijders.1995}. The $i$-th node in the $j$-th module will be simulated as 
\bq {\bf y}_{c(j-1) +i}= {\bf x}_{c(j-1)+1} + N(0,\sigma^2 I_n) \quad \mbox{ for }  \quad 1 \leq i \leq c, \; 1 \leq j \leq k. \label{eq:rg-module} \eq
This introduce a modular structure in the network. We assumed there were total $k=2, 4,5, 10$ modules and each module consists of $c=p/k= 10, 5,4, 2$ nodes. This yields the correlation matrix $C^2=(c_{ij}^2) = (corr({\bf y}_i,{\bf y}_j))$ and the subsequent edge weights  $\sqrt{1- c^2_{ij}}$.

\subsection{No network difference} 
It was expected there was no network difference between networks generated using the same parameters and initial data vectors ${\bf x}_i$ in (\ref{eq:rg-module}). We compared networks with identical number of modules: 4 vs. 4, 5 vs. 5 and 10 vs. 10. It is expected we should not able to detect the network differences.  The performance results were given in terms of the false positive error rate computed as the fraction of  simulations that give $p$-value below 0.05 (Table \ref{table:simulation}). For all the distances except KS-distance, permutation test was used. Since there were 5 samples in each group, the total number of permutations was ${10 \choose 5} = 272$ making the permutation test exact and the comparisons accurate. 
All the distances performed very well. KS-distance was overly sensitive and was producing up to 7\% false positives. However, for 0.05 level test, it is expected there is 5\% chance of producing false positives. Thus, KS-distance is producing only 2\% above the expected error rate. 

\begin{table}[!tb]
\caption{\label{table:simulation}$p=20$ nodes simulation results given in terms of false positive and negative error rates.}
\centering 
\begin{tabular}{|c|ccccccc|} 
\hline
$p$=20 & $L_1$  &    $L_2$ & $L_{\infty}$ & Bottle & GH & KS ($\beta_0$) & KS ($\beta_1$)\\
\hline
    4 vs. 4 &  
    0.00 & 0.00 & 0.00 & 0.00 & 0.00 & 0.04 & 0.01\\
    5 vs. 5 &
    0.00 &
    0.00 &
    0.00 &
    0.00 &
    0.00 &
    0.07 &
    0.01\\
    10 vs. 10 & 0.00 & 0.00 & 0.00 & 0.00 & 0.00 & 0.00 & 0.00 \\
    \hline
    4 vs. 5 &
     0.63&
    0.40 &
    0.33 &
    0.91 &
    0.15 &
    0.27 &
    0.06\\
    2 vs. 4 &
     0.71 &
    0.48 &
    0.42 &
    0.59 &
    0.53 &
    0.18 &
    0.00 \\
    5 vs. 10 &
      0.94 &
    0.80 &
    0.78 &
    0.61 &
    0.72 &
    0.44 &
    0.24\\
\hline
\end{tabular}  
\end{table}

\begin{table}[!tb]
\caption{\label{table:simulation2}$p=100$ nodes simulation results given in terms of false negative error rates.}
\centering 
\begin{tabular}{|c|ccccccc|} 
\hline
$p$=100 & $L_1$  &    $L_2$ & $L_{\infty}$ & Bottle & GH & KS ($\beta_0$) & KS ($\beta_1$)\\
\hline
    4 vs. 5 &
     0.51&
    0.37 &
    0.35 &
    0.83 &
    0.16&
    0.11 &
    0.00\\
    2 vs. 4 &
     0.66 &
    0.45 &
    0.57 &
    0.74 &
    0.61 &
    0.03 &
    0.00 \\
    5 vs. 10 &
     0.94 &
    0.86 &
    0.79 &
    0.74 &
    0.72 &
    0.11 &
    0.00\\
\hline
\end{tabular}  
\end{table}

\subsection{Network differences}  
We generated networks with 2, 4, 5 and 10 modules in the $p=20$ node simulation. Since the number of modules were different, they were considered as different networks. The performance results were given in terms of the false negative error rate computed as the fraction of  simulations that give $p$-value above 0.05 (Table \ref{table:simulation}). All the distance performed badly although KS-distance performed the best. Here, $p=20$ might be too small a network to extract topologically distinct features that are used in topological distances. Thus, we increased number of nodes to $p=100$. All the network distances are still performing badly except KS-distances (Table \ref{table:simulation2}). KS-distance on the number of cycles seem to be the best network distance to use although it has tendency to produce false positives.

In terms of computation, distance methods based on the permutation test took about 950 seconds (16 minutes) while the KS-like test procedure only took about 20 seconds in a computer. The {\tt MATLAB} code for performing these simulations is given in \url{http://www.stat.wisc.edu/~mchung/twins}. The results given in Table \ref{table:simulation} may slightly change if different random networks are generated. 

\section{Application}

\subsection{Dataset and image preprocessing}

\begin{figure}[t]
\centering
\includegraphics[width=1\linewidth]{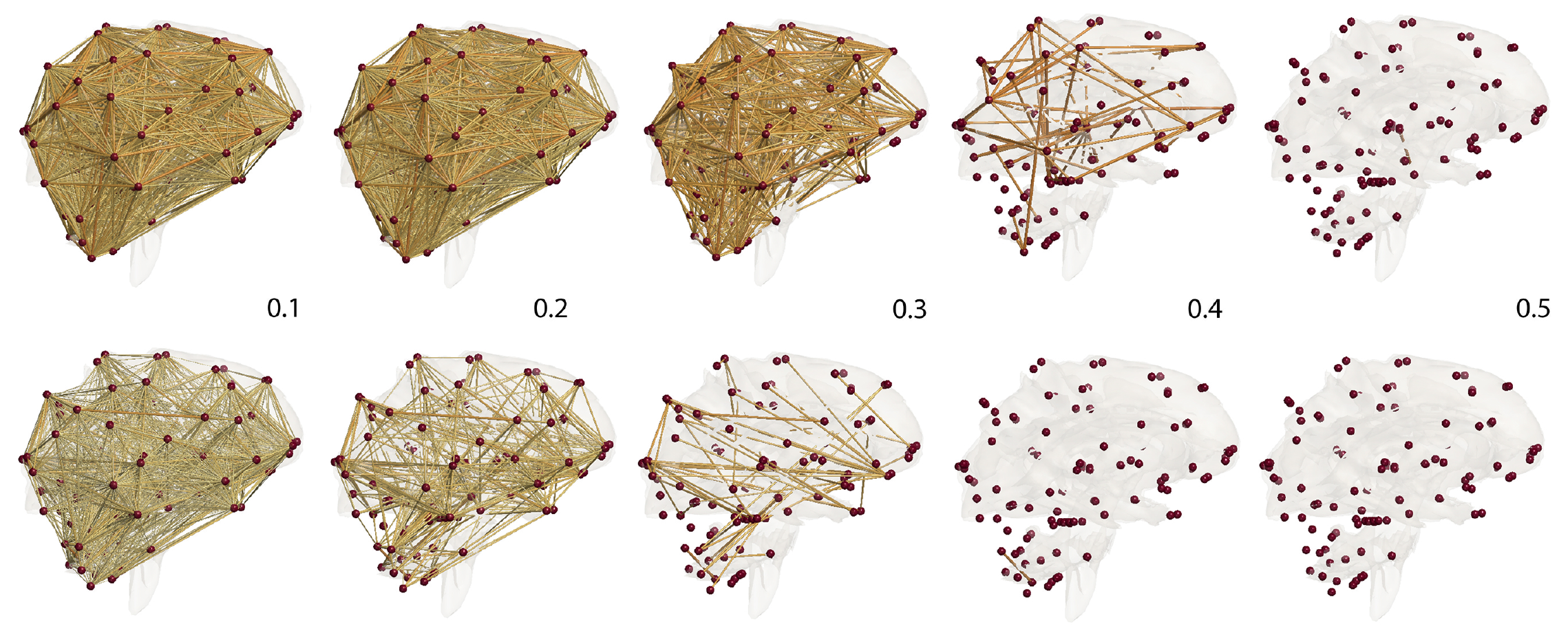}
\caption{Correlation network filtration thresholded at the indicated correlation values. MZ-twins (top) shows more higher correlation connections compared to DZ-twins (bottom). Such connectivity difference is contributed to heritability.}
\label{fig:filtration-correlation}
\end{figure}

We used the resting-state fMRI of 271 twin pairs from the Human Connectome Project \citep{vanessen.2012}. Out of total 271 twin pairs, we only used genetically confirmed 131 monozygotic (MZ) twin pairs (age $29.3\pm3.3$ years, 56M/75F) and 77 same-sex dizygotic (DZ) twin pairs (age $29.1\pm3.5$ years, 30M/47F) in this study.  Since the discrepancy between self-reported and genotype-verified zygosity was fairly hight at 13\% of all the available data, 19 MZ and 19 DZ twin pairs that do not have genotyping were excluded. We additionally excluded 35 twin pairs with missing functional MRI data. 
       
fMRI were collected on a customized Siemens 3T Connectome Skyra scanner, using a gradient-echo-planar imaging (EPI) sequence with multiband factor = 8, TR = 720 ms, TE = 33.1 ms, flip angle = $52^\circ$, $104\times90$ (RO$\times$PE) matrix size, 72 slices, and 2 mm isotropic voxels. 1200 volumes were obtained over 14 min, 33 sec scanning session. fMRI data has undergone spatial and temporal preprocessing including motion and physiological noise removal \citep{smith.2013}.
   Using the resting-state fMRI, we employed the Automated Anatomical Labeling (AAL) brain template to parcellate the brain volume into 116 regions \citep{tzourio.2002}. The fMRI were then averaged across voxels in each brain region for each subject. The averaged fMRI signal in each parcellation is then temporally smoothed using the cosine series representation as follows \citep{chung.2010.SII,gritsenko.2018b}.

Given fMRI time series at the $i$-th parcellation $\zeta_i(t)$ at time $t$, we scaled it to fit to unit interval $[0,1]$. Then subtracted its mean over time $\int_0^1 \zeta_i(t) \;dt$. Then the resulting scaled and translated time series was represented as 
   $$\zeta_i(t) = \sum_{l=0}^k c_{li}\psi_l(t), \; t \in [0,1],$$
 where $\psi_0(t) =1, \psi_l(t) = \sqrt{2} \cos ( l \pi t)$ are cosine basis functions and $c_{li}$ are coefficients estimated in the least squares fashion. For our study, $k=119$ was used such that fMRI were compressed into 10\% of the original data size. $k=119$ expansion increased the signal-to-noise ratio (SNR) as measured by the ratio of variabilities by 81\% in average over all 116 brain regions and 416 subjects, i.e., SNR = 1.81. The resulting real-valued Fourier coefficient vector 
 ${\bf c}_i = (c_{0i}, c_{1i}, \cdots, c_{ki})$ was then used to represent the fMRI in each parcellation as 
 120 features in the spectral domain. 
 
\begin{figure}[t]
\centering
\includegraphics[width=1\linewidth]{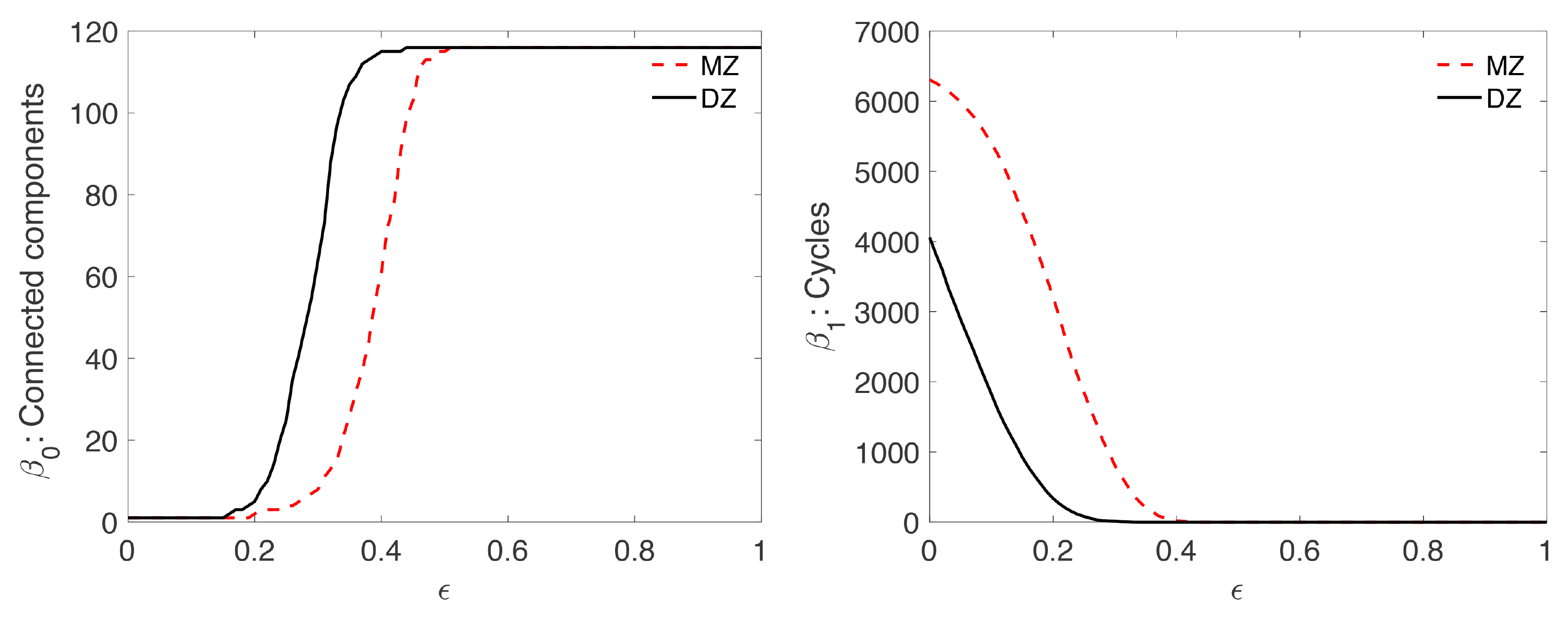}
\caption{Betti-plots showing Betti numbers over correlation $\epsilon$ as filtration. MZ-twins (top) shows more higher correlation connections and cycles compared to DZ-twins (bottom).}
\label{fig:filtration-correlationarrow}
\end{figure}

\subsection{Twin correlations}
The subject level connectivity matrix $C= (c_{ij})$ is computed by correlating 120 features in the spectral domain. Between $i$- and $j$-th parcellations, the connectivity is measured by correlating ${\bf c}_i$ and ${\bf c}_j$ over 120 features, i.e., $c_{ij} = corr({\bf c}_i, {\bf c}_j)$. From the individual  correlation matrices $C$, we computed pairwise twin correlations in each group at the edge level. The resulting group level twin correlations matrices $C_{MZ}=(c^{MZ}_{ij})$ and $C_{DZ}=(c^{DZ}_{ij})$ are nonsymmetric cross-correlation matrices. Since there is no preference in the order of twins, we symmetrize them by $C_{MZ} \leftarrow (C_{MZ} + C_{MZ}^{\top})/2$ and 
$C_{DZ} \leftarrow (C_{DZ} + C_{DZ}^{\top})/2$. The heritability index (HI) is then estimated through  Falconer's formula, which determines the amount of variation due to genetic influence in a population \citep{falconer.1995}:
$$\mbox{HI} = 2 ( C_{MZ} - C_{DZ} ).$$
The network differences between MZ- and DZ- twins are considered as mainly contributed to heritability and can be used to determine the statistical significance of HI \citep{chung.2017.IPMI,chung.2018.MICCAI}. Five network distances were compared except for the KS-distance. All the distance except KS-distance is not practical in this setting due to the large number of permutations and their performance were suboptimal in the simulation study.

In most brain imaging studies, 5000-1000000 permutations are often used, which puts the total number of generated permutations to usually less than 0.01 to 1\% of all possible permutations. In \citet{zalesky.2010}, 5000 permutations are out of a possible ${27 \choose 12}=17383860$ permutations (2.9\%) were used. In \citet{thompson.2001}, for instance, 1 million permutations out of ${40 \choose 20}$ possible permutations (0.07\%) were generated using a super computer. In \citet{lee.2017.HBM}, 5000 permutations out of possible ${33 \choose 10}=92561040$ permutations (0.005\%) were used.
Since we have 131 MZ and 77 DZ pairs, the total number of possible permutation is ${271 \choose 131}$, which is larger than $10^{80}$. Even if we generate only 0.01\% of $10^{80}$ of all possible permutations, $10^{76}$ permutations are still too large for most computers. Thus, we choose the KS-distance for measuring the network distance. Although the probability distribution of the KS-distance is actually based on the permutation test but the probability is computed combinatorially bypassing the need for resampling. KS-distance in our study only took few seconds to compute the $p$-value.

\subsection{Results}
We used $\beta_0$ and $\beta_1$ in computing KS-distances. Denote the element-wise application of  an arbitrary monotone function $\phi$ to matrix $C_{MZ}$ as  $\phi \circ C_{MZ} = (\phi(c_{ij}^{MZ}))$. Then KS-distance between $C_{MZ}$ and $C_{DZ}$ is equivalent to KS-distance between $1-C_{MZ}$ and $1-C_{DZ}$ as well as between $\phi \circ (1-C_{MZ})$ and $\phi \circ (1-C_{DZ})$. Thus, we simply built filtrations over $C_{MZ}$ and $C_{DZ}$ and computed KS-distance without making them into metric. We used 101 filtration values between 0 and 1 at 0.01 increment (Figure \ref{fig:filtration-correlation}). This gives a reasonably accurate estimate of the maximum gap in the $\beta_i$-plots between the twins (Figure \ref{fig:filtration-correlationarrow}). For $\beta_0$-plots, the maximum gap is 82, which gives the $p$-value smaller than $10^{-24}$. For $\beta_1$-plots, the maximum gap is 3647, which gives the $p$-value smaller than $10^{-32}$. At the same correlation value, MZ-twins are more connected than DZ-twins. Also MZ-twins have more cycles than DZ-twins. Such huge topological differences are contributed to heritability. 

\begin{figure}[t]
\centering
\includegraphics[width=1\linewidth]{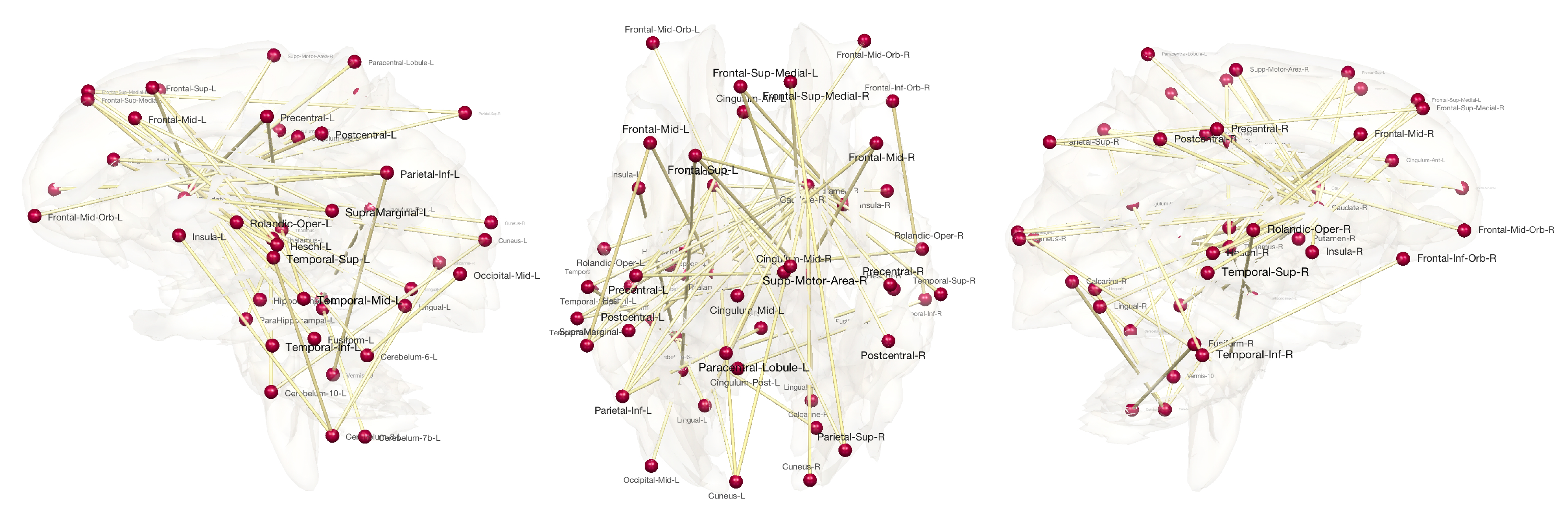}
\caption{Most highly heritable connections. The connections with 100\% heritability are only shown.}
\label{fig:filtration-HI}
\end{figure}

Figure \ref{fig:filtration-HI} displays the HI index that gives 100\% heritability. The most heritable connections include the left frontal gyrus, left and right middle frontal gyri, left superior frontal gyrus, left parahippocampal gyrus, left and right thalami, left and right caudate nuclei among other regions. Most regions overlap with highly heritable regions observed in resting-state connectivity of twins in a different study \citep{glahn.2010}. Moreover, the findings here are consistent with a previous study on diffusion tensor imaging on twins (Chung et al., 2018). The left and right caudate nuclei are identified as most heritable hub nodes. 

\section{Discussion}
\noindent{\em Log-Euclidean distance.}  Another often used matrix-norm based distance is log-Euclidean on positive definite symmetric matrices (PDS).  Given edge weights $ \rho^1, \rho^2$ that are PDS, the log-Euclidean distance between $\rho^1$ and $\rho^2$ is given by \citep{arsigny.2005}
$$d_{LE} (\rho^1, \rho^2) =  \left[ \mbox{tr} \left(  \log (\rho^1) - \log (\rho^2)   \right)^2 \right]^{1/2},$$
where $\log \cdot$ is the matrix logarithm. The log-Euclidean distance can be viewed as the generalized manifold version of Frobenius norm distance.

If a matrix is nonnegative definite with zero eigenvalues, the matrix logarithm is {\em not} defined since $\log 0$ is not defined. Thus, we cannot apply the logarithm directly to  rank-deficient large correlation and covariance matrices obtained from data with small sample sizes relative to the number of nodes.  One way of applying the logarithm to nonnegative definite matrices is to make the matrix diagonally dominant by adding a diagonal matrix $\alpha I$ with suitable choice of relatively large $\alpha$ \citep{chan.1997}. Alternately, we can perform a graphical LASSO-type of sparse model and obtain the closest positive definite matrices \citep{qiu.2015,mazumder.2012}.  Developing the log-Euclidean for general correlation matrices including nonnegative ones is beyond the scope of this paper. This is left as a future study. Note topological distances are  applicable to nonnegative definite connectivity matrices.

\begin{wrapfigure}{r}{0.5\textwidth}
\centering
\includegraphics[width=1\linewidth]{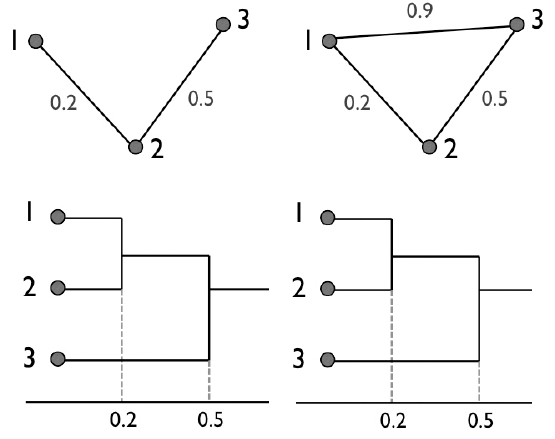}
\caption{Two topologically distinct graphs may have identical dendrograms, which results in zero GH-distance.} 
\label{fig:topology-dendro2}
\vspace{-0.3cm}
\end{wrapfigure}

\noindent{\em The limitation of GH-distances.}  The limitation of the existing topological distances such as GH-distance is the inability to discriminate cycles in a graph. Consider two topologically different graphs with three nodes (Figure \ref{fig:topology-dendro2}). 
However, the corresponding single linkage matrix (SLM) are identically given by
\bqn
 \left( \begin{array}{ccc}
0 & 0.2 & 0.5\\ 
0.2 & 0 & 0.5\\
 0.5    &0.5  & 0
\end{array} \right) \mbox{ and }  \left( \begin{array}{ccc}
0 & 0.2 & 0.5\\ 
0.2 & 0 & 0.5\\
 0.5    &0.5  & 0
\end{array} \right).\eqn
The lack of uniqueness of SLMs makes GH-distance incapable of discriminating networks with cycles \citep{chung.2017.IPMI}. 
KS-distance also treats the two networks in Figure \ref{fig:topology-dendro2} as identical if  Betti number $\beta_0$ is used as the monotonic feature function. It is possible to develop network distances that are sensitive to presence of cycles using $\beta_1$. Further it may be possible to construct a distance that uses the combination of both $\beta_0$ and $\beta_1$. This is beyond the scope of this paper and left as a future study.

\noindent{\em Computational issues.} The total number of permutations in permuting two groups of size $q$ each is ${2q \choose q} \sim \frac{4^q}{\sqrt{2\pi q}}.$
Even for small $q=10$, more than tens of thousands permutations are needed for the accurate approximation the $p$-value. The main advantage of KS-distance over all other distance measures is that it avoids numerically performing the permutation test and avoids generating tens of thousand permutations. Although the probability distribution of the KS-distance is actually based on the permutation test but the probability is computed combinatorially. We believe it is possible to develop similar theoretical results for other distance measures and come up with a method for avoiding resampling based method for statistical inference. This is left as a future study.

\section*{Acknowledgements}
This study was supported by NIH Grant 
UL1TR000427, Brain Initiative Grant EB022856 and Basic Science Research Program through the National Research Foundation (NRF) of Korea (NRF-2016R1D1A1B03935463). We would like to thank Zhiwei Ma of University of Chicago for providing Example 2, Yuan Wang of University of South Carolina, Alex Leow of University of Illinois-Chicago, Guorong Wu of University of North Carolina and Martin Lindquist of Johns Hopkins University for valuable discussions and Gregory Kirk of University of Wisconsin for the logistic support  on HCP data.

\bibliographystyle{abbrvnat}
\bibliography{reference.2018.08.01}

\end{document}